\documentclass[a4paper,11pt]{article}

\usepackage{jheppub} 
\usepackage{graphicx}

\usepackage[T1]{fontenc} 

\newcommand{\tr}{{\rm tr}}
\newcommand{\im}{\mathop{\rm Im}}
\newcommand{\re}{\mathop{\rm Re}}

\newcommand{\CA}{{\cal A}}
\newcommand{\CH}{{\cal H}}

\newcommand{\CN}{{\cal N}}
\newcommand{\CO}{{\cal O}}
\newcommand{\CS}{{\cal S}}
\newcommand{\CT}{{\cal T}}

\newcommand{\vk}{{\vec k}\hspace{.5pt}}
\newcommand{\vl}{{\vec l}\hspace{2.5pt}}
\newcommand{\vp}{{\vec p}\hspace{1pt}}
\newcommand{\vq}{{\vec q}\hspace{1pt}}
\newcommand{\vx}{{\vec x}}

\title{\boldmath Variation of Entanglement Entropy in Scattering Process}

\author[a]{I.Y.~Park,}
\author[b]{Shigenori Seki}
\author[c]{and Sang-Jin Sin}

\affiliation[a]{Department of Applied Mathematics, Philander Smith College\\Little Rock, AR 72223, USA}
\affiliation[b]{Research Institute for Natural Science, Hanyang University\\Seoul 133-791, Republic of Korea}
\affiliation[c]{Department of Physics, Hanyang University,\\Seoul 133-791, Republic of Korea}

\emailAdd{inyongpark05@gmail.com}
\emailAdd{sigenori@hanyang.ac.kr}
\emailAdd{sjsin@hanyang.ac.kr}

\abstract{In a scattering process, the final state is determined by an initial state and an S-matrix. 
We focus on two-particle scattering processes and consider the entanglement between these particles. 
For two  types  initial states; {\it i.e.}, 
an unentangled state and an entangled one, 
we  calculate perturbatively the change of entanglement entropy 
from the initial state to the final one. 
Then we show a few examples in a field theory and in quantum mechanics. 
}

\begin{document} 
\maketitle
\flushbottom

\section{Introduction}
\label{sec:intro}

Entanglement is a characteristic feature in a quantum theory. 
The entanglement in quantum field theories has been studied extensively in the past decade. 
When one considers a sub-system $A$ and its complement $\overline{A}$, 
the entanglement entropy between $A$ and $\overline{A}$ 
is defined by the von Neumann entropy $S_E = -\tr_A \rho_A \log \rho_A$ 
with the reduced density matrix $\rho_A$. 
Calabrese and Cardy have systematically studied it in a conformal field theory 
with the use of a replica trick \cite{Calabrese:2004eu}. 
The other remarkable recent progress is 
the holographic derivation of entanglement entropy 
suggested by Ryu and Takayanagi \cite{Ryu:2006bv,Ryu:2006ef}. 
Following it, one can obtain an entanglement entropy by calculating $S_E = \CA/(4 G_N)$, 
where $\CA$ is the area of a minimal surface whose boundary is the boundary of the sub-system A. 
In other words, the holographic entanglement entropy provides us with a geometric understanding 
of entanglement. 

Then there is the other geometric interpretation of entanglement entropy 
conjectured recently by Maldacena and Susskind \cite{Maldacena:2013xja}. 
Its original purpose was to resolve the firewall paradox \cite{Almheiri:2012rt}\footnote{See 
Ref.~\cite{Braunstein:2009my} for an earlier work. 
It has predicted an energetic curtain, which is similar to the firewall, 
on the assumptions different from Ref.~\cite{Almheiri:2012rt}.}. 
This conjecture states that an Einstein-Rosen-Podolski pair, {\it i.e.}, a pair of entangled objects, 
is connected by an Einstein-Rosen bridge (or a wormhole). 
Therefore the conjecture is symbolically called the ER=EPR conjecture. 
From the point of view of the AdS/CFT correspondence, 
some examples supporting the ER=EPR conjecture have been shown. 
An entangled pair of accelerating quark and anti-quark was studied in Ref.~\cite{Jensen:2013ora}. 
Investigating the causal structure on the world-sheet minimal surface 
that is the holographic bulk dual of such a quark and anti-quark on the AdS boundary, 
Ref.~\cite{Jensen:2013ora} has found that there exists a wormhole on the minimal surface
and that any open strings connecting the quark and anti-quark must go through the wormhole. 
Therefore the entanglement of the accelerating quark and anti-quark 
coincides with the existence of the wormhole. 
Furthermore, Ref.~\cite{Sonner:2013mba} considered Schwinger pair creation of 
a quark and an anti-quark and confirmed that there is a wormhole on the string world-sheet 
of their bulk dual. 
Ref.~\cite{Seki:2014pca} focused on a pair of scattering gluons as an EPR pair. 
Since Ref.~\cite{Alday:2007hr} had shown the minimal surface solution 
corresponding to the gluon scattering, 
Ref.~\cite{Seki:2014pca} calculated the induced metric on the minimal surface 
and found a wormhole connecting the gluon pair. 
One can then naturally guess that, in a scattering process\footnote{Ref.~\cite{Lello:2013bva} has 
studied the entanglement entropy in a decay process in terms of the Wigner-Weisskopf method.}, 
an interaction induces the variation of entanglement 
from an incoming state to an outgoing one.
We know these states are associated with each other by an S-matrix. 
So the question is how the variation of entanglement entropy and the S-matrix 
are related. 
In this paper we attack this problem by a perturbative analysis in a weak coupling $\lambda$. 
In order to evaluate the entanglement entropy, 
it is useful to calculate R{\' e}nyi entropy by the replica trick when one can calculate it exactly. 
For instance, Ref.~\cite{Mollabashi:2014qfa} explicitly calculated 
the time evolution of the entanglement entropy between two free scalar field theories 
with a specific interaction. 
However, this method is often unavailable for a perturbative analysis. 
Therefore we apply the method developed by Ref.~\cite{Balasubramanian:2011wt,Hsu:2012gk}, in which 
the entanglement between two divided momentum spaces was studied perturbatively.

In Section 2, we consider the variations of entanglement entropy from two kinds of initial states; 
one is an unentangled initial state and the other is an entangled one. 
In Section 3, we evaluate the variation of entanglement entropy 
in the field theory with a $\phi^4$-like interaction. 
We also consider the time-dependent interaction in quantum mechanics. 
Section 4 is devoted to conclusion and discussion.

\section{Perturbative calculation of entanglement entropy}
\label{sec:pertu}

Since we are interested in a scattering process of two particles, A and B, and their entanglement, 
let us consider the Hamiltonian with an interaction: 
\begin{align}
H = H_0 + \lambda H_{\rm int} \,, \quad H_0 = H_A \otimes {\bf 1} + {\bf 1} \otimes H_B \,. \label{eq:Hami}
\end{align} 
It is usually difficult to divide the total Hilbert space $\CH$ to $\CH_A \otimes \CH_B$ 
due to the interaction. 
However an initial state far in the past and a final state far in the future in a scattering process 
can be regarded as states generated by an asymptotically free Hamiltonian. 
Furthermore, although a field theory in general includes arbitrary multi-particle states 
in its Hilbert space, 
we concentrate only on an elastic scattering of two particles 
such as ${\rm A}+ {\rm B} \to {\rm A}+ {\rm B}$ with a weak coupling. 
That is to say, we restrict the Hilbert space to the (1+1)-particle Fock space, 
in which the initial and final states are. 
Since such a restriction usually violates unitarity for local interaction terms, 
we assume in this paper specific theories that do not produce 
states of more than 1+1 particles at lower orders of perturbation 
(see an example in Section \ref{sec:phi4}). 
Then the unitarity is approximately recovered at a weak coupling. 
Under this assumption, we can divide the Hilbert space of the initial and final states to $\CH_A \otimes \CH_B$, 
and these states are denoted by a (1+1)-particle state generated by the free Hamiltonian $H_0$, 
namely, a state of a particle A and B with momentum $p$ and $q$: 
\begin{align}
|p,q\rangle := |p\rangle_A \otimes |q\rangle_B \,.
\end{align}
One can express the infinite time evolution from the initial state to the final one 
in terms of S-matrix by definition,  
\begin{align}
\lim_{t\to\infty}\langle{\rm fin}|e^{-iHt} |{\rm ini}\rangle 
= \langle{\rm fin}|{\bf S}|{\rm ini}\rangle \,, \quad 
{\bf S} := {\bf 1} + i{\bf T} \,.
\end{align}
${\bf T}$ is a transition matrix in $\CO(\lambda)$ which is induced by the interaction. 
Then the final state is described as 
\begin{align}
|{\rm fin}\rangle = \int dk dl\, |k,l\rangle \langle k,l|{\bf S}|{\rm ini}\rangle \,, \label{eq:finsta}
\end{align}
in which we used the completeness relation of (1+1)-particles' states, {\it i.e.}, 
$({\bf 1})_\text{(1+1)-particles} = \int dkdl\, |k,l\rangle \langle k,l|$, 
and an inner product of states, {\it i.e.}, $\langle k,l|p,q\rangle = \delta(k-p)\delta(l-q)$. 
Although the norm $\langle p,q|p,q\rangle =:V$ has an infinite volume, 
we shall fix a normalization at the stage of a reduced density matrix. 
Here we comment that one can easily formulate the case of discrete spectra 
by replacing $\int dk dl$ with $\sum_{k,l}$. 
As an example we shall show in Section \ref{sec:timedep} 
the theory with a time-dependent interaction in non-relativistic quantum mechanics.

The total density matrix of the final state 
is $\rho^{\rm (fin)} = |{\rm fin}\rangle \langle {\rm fin}|$, 
and we obtain the reduced density matrix $\rho_A^{\rm (fin)}$ 
by taking trace of $\rho^{\rm (fin)}$ with respect to the particle B, {\it i.e.}, 
$\rho_A^{\rm (fin)} = \tr_B \rho^{\rm (fin)}$ up to normalization. 
In the case of (\ref{eq:finsta}) we can write down the reduced density matrix as 
\begin{align}
\rho_A^{\rm (fin)} = {1 \over \CN}\int dk dk' \biggl( \int dl \langle k,l|{\bf S}|{\rm ini}\rangle \langle {\rm ini}|{\bf S}^\dagger|k',l\rangle \biggr) |k\rangle \langle k'| \,, \label{eq:redDM}
\end{align}
where $\CN$ is a normalization constant determined by $\tr_A \rho_A^{\rm (fin)} = 1$, namely, 
\begin{align}
\CN = \int dkdl\, |\langle k,l|{\bf S}|{\rm ini}\rangle|^2 \,. \label{Eq:genenorm}
\end{align}
Then the entanglement entropy between A and B in the final state is 
\begin{align}
S_E^{\rm (fin)} = -\tr \rho_A^{\rm (fin)} \log \rho_A^{\rm (fin)} \,, \label{eq:entent}
\end{align}
and the variation of entanglement entropy from the initial state to the final one is 
\begin{align}
\Delta S_E = S_E^{\rm (fin)} -S_E^{\rm (ini)} \,, \label{eq:EEchange}
\end{align}
where $S_E^{\rm (ini)}$ is the entanglement entropy of the initial state. 
We shall calculate these entanglement entropies perturbatively.

The replica trick allows us to calculate a R{\' e}nyi entropy, 
$S(n) = {1 \over 1-n}\log \tr_A \rho_A^n$. 
The entanglement entropy is given by the $n \to 1$ limit of R{\' e}nyi entropy, namely, 
$S_E = \lim_{n\to 1}S(n) = -\lim_{n\to 1}{\partial \over \partial n}\tr_A \rho_A^n$. 
Therefore the method to derive an entanglement entropy via a R{\' e}nyi entropy 
is often useful. 
However, we are confronted with a problem 
when we analyze a quantum theory with a coupling $\lambda$ in terms of perturbation. 
When one obtains a perturbative expansion of $\tr_A\rho_A^n$, 
the term of order $\lambda^n$ relevantly contributes to the entanglement entropy 
because the operation $\lim_{n\to 1}{\partial \over \partial n}$ acts on $\lambda^n$ 
and yields a term of $\lambda\log\lambda$ order. 
In other words, the higher order terms in the R{\' e}nyi entropy are responsible 
for the convergence of the entanglement entropy under the $n \to 1$ limit. 
Hence any $\lambda^n$-order terms in $\tr_A\rho_A^n$ are necessary 
in order to obtain a meaningful entanglement entropy. 
In this paper, instead of the replica trick, 
we apply the perturbative method developed by Ref.~\cite{Balasubramanian:2011wt} 
for calculating an entanglement entropy.

\subsection{Unentangled initial state}
\label{subsec:unent}

Let us consider the simplest single state with fixed momenta $p_1$ and $q_1$ for the initial state of particle A and B, 
\begin{align}
|{\rm ini}\rangle \sim |p_1,q_1\rangle \,. \label{eq:uneini}
\end{align}
The normalization of states will be properly fixed later 
in normalizing a density matrix so that $\tr_A \rho_A^{\rm (fin)} = 1$. 
This initial state is obviously unentangled, {\it i.e.}, $S_E^{\rm (ini)} = 0$. 
Then we can describe the final state (\ref{eq:finsta}) as 
\begin{align}
|{\rm fin}\rangle &= \int dk dl\, |k,l\rangle \CS_{kl;p_1q_1} \nonumber\\
&= {\CS_{p_1q_1;p_1q_1} \over V^2}|p_1,q_1\rangle 
	+ i\lambda\int_{k\neq p_1}dk\, {\CT_{kq_1;p_1q_1} \over V} |k,q_1\rangle 
	+ i\lambda\int_{l\neq q_1} dl\, {\CT_{p_1l;p_1q_1} \over V} |p_1,l\rangle \nonumber\\
&\quad	+ i\lambda\int_{k\neq p_1 \atop l \neq q_1}dkdl\, \CT_{kl;p_1q_1} |k,l\rangle \,, \label{eq:unefin} 
\end{align}
where we introduced an infinite spacial volume $V := \int dx\, e^{ix\cdot 0} = \delta(0)$ 
due to the divergence of norms, {\it i.e.}, $\langle p |p \rangle_A = \langle q |q \rangle_B = \delta(0)$.
The integral $\int_{k\neq p}dk$ means $\int dk (1-V^{-1}\delta(k-p))$. 
$\CS_{kl;pq}$ and $\CT_{kl;pq}$ denote S and T-matrix elements, 
\begin{align}
\CS_{kl;pq} := \langle k,l|{\bf S}|p,q\rangle \,, \quad 
\CT_{kl;pq} := {1 \over \lambda}\langle k,l|{\bf T}|p,q\rangle \,. \label{eq:CSCTdef}
\end{align}
${\bf S}$ includes an identity ${\bf 1}$, 
while ${\bf T}$ is given by an interaction with coupling $\lambda$. 
Therefore the possible lowest orders of (\ref{eq:CSCTdef}) 
with respect to $\lambda$ are 
\begin{align}
\CS_{pq;pq} \sim \CO(\lambda^0) \,, \quad 
\CS_{p'q';pq}|_{(p',q')\neq (p,q)} \sim \CO(\lambda) \,, \quad 
\CT_{kl;pq} \sim \CO(\lambda^0) \,. \label{eq:CSCTorder}
\end{align}
We employ the method developed by Ref.~\cite{Balasubramanian:2011wt} 
in order to perturbatively calculate the entanglement entropy. 
Since Eq.~(\ref{eq:unefin}) is rewritten as 
\begin{align}
|{\rm fin}\rangle = {\CS_{p_1q_1;p_1q_1} \over V^2} |{\tilde p}_1\rangle_A \otimes |{\tilde q}_1\rangle_B 
	+\int_{k\neq p_1 \atop l \neq q_1}dkdl \biggl( \lambda^2{\CT_{kq_1;p_1q_1} \CT_{p_1 l;p_1q_1} \over \CS_{p_1q_1;p_1q_1}} +i\lambda \CT_{kl;p_1q_1}\biggr) |k,l\rangle \,, 
\end{align}
with 
\begin{align}
|{\tilde p}_1\rangle_A = |p_1\rangle_A + i \lambda V \int_{k\neq p_1} dk{\CT_{kq_1;p_1q_1} \over \CS_{p_1q_1;p_1q_1}}|k\rangle_A \,, \quad 
|{\tilde q}_1\rangle_B = |q_1\rangle_B + i \lambda V \int_{l\neq q_1} dl{\CT_{p_1 l;p_1q_1} \over \CS_{p_1q_1;p_1q_1}}|l\rangle_B \,,  
\end{align}
we can calculate the reduced density matrix (\ref{eq:redDM}) as 
\begin{align}
\rho_A^{\rm (fin)} &= {1 \over \CN_1} \biggl( 
	{|\CS_{p_1q_1;p_1q_1}|^2 \over V^3} |{\tilde p}_1\rangle\langle {\tilde p}_1| 
	+\lambda^2 V^2 \int_{k,k'\neq p_1}dk dk' M_{kk'} |k\rangle \langle k'| \biggr) \,,  \nonumber\\
M_{kk'} &= {1 \over V^2} \int_{l\neq q_1} dl
	\biggl( \lambda{\CT_{kq_1;p_1q_1} \CT_{p_1 l;p_1q_1} \over \CS_{p_1q_1;p_1q_1}} +i\CT_{kl;p_1q_1}\biggr)
	\biggl( \lambda{\CT_{k'q_1;p_1q_1} \CT_{p_1 l;p_1q_1} \over \CS_{p_1q_1;p_1q_1}} +i\CT_{k'l;p_1q_1}\biggr)^* \,. \nonumber\\ \label{eq:uenredDM}
\end{align}
$\CN_1$ is the normalization factor which is fixed by $\tr_A \rho_A^{\rm (fin)} = 1$, namely, 
\begin{align}
\CN_1 = {|\CS_{p_1q_1;p_1q_1}|^2 \over V^2} + \lambda^2 V^2 \int_{k\neq p_1}dk\, M_{kk} \,. 
\end{align}
Here we recall (\ref{eq:CSCTorder}) and it leads to $M_{kk'} \sim \CO(1)$.  
After a perturbative expansion, the reduced density matrix (\ref{eq:uenredDM}) becomes 
\begin{align}
\rho_A^{\rm (fin)} &= 
	\biggl(1-\lambda^2\int_{k\neq p_1} dk\, M_{kk} \biggr){1 \over V}|{\tilde p}_1\rangle \langle {\tilde p}_1| 
	+\lambda^2 \int_{k,k'\neq p_1}dk dk'\, M_{kk'} |k\rangle \langle k'| 
	+\CO(\lambda^3) \,, \\
M_{kk'} &= {1 \over V^2}\int_{l\neq q_1} dl\, \CT_{kl;p_1q_1}\CT_{k'l;p_1q_1}^* +\CO(\lambda) \,, \quad (k,k'\neq p_1)
\end{align}
When the eigenvalues of $M_{kk'}$ at leading order 
are denoted by $m_k$, we obtain 
\begin{align}
\int_{k\neq p_1}dk\, M_{kk} = \tr_A M_{kk'} 
= \int_{k\neq p_1}dk\, m_k \,,
\end{align}
up to $\CO(\lambda)$. 
Therefore the entanglement entropy of final state (\ref{eq:entent}) becomes 
\begin{align}
S_E^{\rm (fin)} &= -\biggl(1 -\lambda^2\int_{k\neq p_1}dk\, m_k \biggr) \log \biggl(1 -\lambda^2\int_{k\neq p_1}dk\, m_k \biggr) \nonumber\\ 
&\quad - \int_{k\neq p_1}dk\, (\lambda^2 m_k) \log (\lambda^2 m_k) + \CO(\lambda^3) \nonumber\\ 
&= -\lambda^2\log\lambda^2 \int_{k\neq p_1}dk\, m_k 
	+\lambda^2 \int_{k\neq p_1}dk\, m_k (1 -\log m_k) 
	+ \CO(\lambda^3) \,. \label{eq:uneEEchange}
\end{align}
Only the T-matrix elements $\CT_{kl;p_1q_1}$ with $k \neq p_1 $ and $l \neq q_1$ 
contribute to the entanglement entropy of the final state at leading order. 
Of course, since the entanglement entropy of the unentangled initial state (\ref{eq:uneini}) vanishes, 
the variation of entanglement entropy (\ref{eq:EEchange}), $\Delta S_E$, 
is equal to $S_E^{\rm (fin)}$ itself in (\ref{eq:uneEEchange}).

\subsection{Entangled initial state}

Let us consider an entangled initial state, 
\begin{align}
|{\rm ini}\rangle \sim u_1|p_1,q_1\rangle +u_2|p_2,q_2\rangle \,, \label{eq:entini}
\end{align}
with $p_1\neq q_1$, $p_2 \neq q_2$, $u_1^2 + u_2^2 = 1$, $u_{1,2} \neq 0$ and $u_1 \geq u_2$.
The entanglement entropy of this state is 
\begin{align}
S_E^{\rm (ini)} = \sum_{j = 1}^2 |u_j|^2 \log |u_j|^2 \,. \label{eq:entinient}
\end{align}

We can write down the final state in terms of the S-matrix (or T-matrix), 
\begin{align}
|{\rm fin}\rangle 
&= {\CS_{p_1q_1} \over V^2}|p_1,q_1\rangle 
	+ {\CS_{p_2q_2} \over V^2}|p_2,q_2\rangle 
	+i\lambda {\CT_{p_1q_2} \over V^2} |p_1,q_2\rangle 
	+i\lambda {\CT_{p_2q_1} \over V^2}|p_2,q_1\rangle  \nonumber\\ 
&\quad	+i\lambda\int_{l\neq q_1,q_2} dl\, \sum_{j=1}^2 {\CT_{p_j l} \over V} |p_j,l\rangle 
	+i\lambda\int_{k\neq p_1,p_2} dk\, \sum_{j=1}^2 {\CT_{k q_j} \over V} |k, q_j\rangle \nonumber\\
&\quad	+i\lambda\int_{k\neq p_1,p_2 \atop l \neq q_1,q_2} dkdl\, \CT_{kl}|k,l\rangle \,, \label{eq:unent12}
\end{align}
where 
\begin{align}
\CS_{kl} := u_1 \CS_{kl;p_1q_1} +u_2 \CS_{kl;p_2q_2} \,, \quad
\CT_{kl} := u_1 \CT_{kl;p_1q_1} +u_2 \CT_{kl;p_2q_2}\,. 
\end{align}
Note that $\CS_{p_1q_1} = u_1 V^2  + i\lambda \CT_{p_1q_1}$ and 
$\CS_{p_2q_2} = u_2 V^2 + i\lambda \CT_{p_2q_2}$. 
Firstly we diagonalize the first line in (\ref{eq:unent12}) by the use of 
\begin{align}
\mathbb{Q} = \begin{pmatrix} \CS_{p_1q_1} & i\lambda\CT_{p_1q_2} \\ i\lambda \CT_{p_2q_1} &\CS_{p_2q_2}\end{pmatrix} \,, \quad 
W = \begin{pmatrix} 
		i\lambda\CT_{p_2q_1} & \CS_{p_2q_2} -\zeta_2 \\ 
		\CS_{p_1q_1} -\zeta_1 & i\lambda\CT_{p_1q_2} 
	\end{pmatrix} \,,\quad
W \mathbb{Q} W^{-1} = 
	\begin{pmatrix} 
		\zeta_1 & 0 \\ 
		0 &\zeta_2 
	\end{pmatrix} \,, 
\end{align}
where
\begin{align}
\zeta_1 +\zeta_2 = \CS_{p_1q_1}+\CS_{p_2q_2} \,, \quad 
\zeta_1 -\zeta_2 = \sqrt{(\CS_{p_1q_1}-\CS_{p_2q_2})^2 -4\lambda^2\CT_{p_1q_2}\CT_{p_2q_1}} \,.
\end{align}
Following this diagonalization, the basis is transformed as 
\begin{align}
\begin{pmatrix} |p_1\rangle \\ |p_2\rangle \end{pmatrix} 
	= W^t \begin{pmatrix} |{\hat p}_1\rangle \\ |{\hat p}_2\rangle \end{pmatrix} \,, \quad
\begin{pmatrix} |q_1\rangle \\ |q_2\rangle \end{pmatrix}
	= W^{-1} \begin{pmatrix} |{\hat q}_1\rangle \\ |{\hat q}_2\rangle \end{pmatrix} \,. \label{eq:diagvec}
\end{align}
Then we can rewrite the final state (\ref{eq:unent12}) as
\begin{align}
|{\rm fin}\rangle 
&= \sum_{j=1}^2{\zeta_j \over V^2} |{\hat p}_j,{\hat q}_j\rangle 
	+i\lambda\int_{l\neq q_1,q_2} dl\, \sum_{j=1}^2 {A_j(l) \over V} |{\hat p}_j,l\rangle 	+i\lambda\int_{k\neq p_1,p_2} dk\, \sum_{j=1}^2 {B_j(k) \over V} |k,{\hat q}_j\rangle \nonumber\\ 
&\quad	+i\lambda\int_{k\neq p_1,p_2 \atop l \neq q_1,q_2} dkdl\, \CT_{kl}|k,l\rangle \,,
\end{align}
where 
\begin{align}
&A_1(l) = i\lambda \CT_{p_1l}\CT_{p_2q_1} +\CT_{p_2l}(\CS_{p_2q_2}-\zeta_2) \,, \quad 
A_2(l) = \CT_{p_1l}(\CS_{p_1q_1}-\zeta_1) +i\lambda\CT_{p_2l}\CT_{p_1q_2} \,, \nonumber\\
&B_1(k) = {i\lambda\CT_{kq_1}\CT_{p_1q_2} -\CT_{kq_2}(S_{p_1q_1}-\zeta_1) \over \det W} \,, \quad
B_2(k) = {-\CT_{kq_1}(S_{p_2q_2}-\zeta_2) +i\lambda\CT_{kq_2}\CT_{p_2q_1}  \over \det W} \,.
\end{align}
Furthermore we can rearrange the basis so that 
\begin{align}
|{\rm fin}\rangle = \sum_{j=1}^2 {\zeta_j \over V^2} |{\tilde p}_j\rangle_A \otimes |{\tilde q}_j\rangle_B 
	+\int_{k\neq p_1,p_2 \atop l \neq q_1,q_2} dkdl\, \Biggl(\lambda^2\sum_{j=1}^2{A_j(l)B_j(k) \over \zeta_j} +i\lambda\CT_{kl}\Biggr)|k,l\rangle \,,
\end{align}
where 
\begin{align}
|{\tilde p}_j\rangle_A &= |{\hat p}_j\rangle_A +i\lambda {V \over \zeta_j} \int_{k\neq p_1,p_2} dk\, B_j(k)|k\rangle_A \,, \nonumber\\
|{\tilde q}_j\rangle_B &= |{\hat q}_j\rangle_B +i\lambda {V \over \zeta_j} \int_{l\neq q_1,q_2} dk\, A_j(k)|k\rangle_B \,. \quad (j=1,2) \label{eq:defredvec}
\end{align}
As a result, we obtain the reduced density matrix (\ref{eq:redDM}) after a similarity transformation,  
\begin{align}
\rho_A^{\rm (fin)} &= {1 \over \CN_2}\Biggl( 
	\sum_{j=1}^2 {|\zeta_j|^2 \over V^3} |{\tilde p}_j\rangle\langle {\tilde p}_j| 
	+\lambda^2 V^2 \int_{k,k'\neq p_1,p_2}dk dk' R_{kk'} |k\rangle \langle k'| \Biggr) \,, \\
R_{kk'} &= {1 \over V^2} \int_{l\neq q_1,q_2} dl
	\Biggl(\lambda\sum_{j=1}^2{A_j(l)B_j(k) \over \zeta_j} +i\CT_{kl}\Biggr)
	\Biggl(\lambda\sum_{j=1}^2{A_j(l)B_j(k') \over \zeta_j} +i\CT_{k'l}\Biggr)^* \,. \nonumber\\ \label{eq:entredDM}
\end{align}
The leading term of $R_{kk'}$ does not depend on $A_j$ and $B_j$ 
but on $\CT_{kl}$ ($k\neq p_1,p_2$, $l\neq q_1,q_2$). 
Using the normalization $\tr_A\rho_A^{\rm (fin)} = 1$, $\CN_2$ is computed as 
\begin{align}
\CN_2 = \sum_{i=1}^2 {|\zeta_j|^2 \over V^2} + \lambda^2 V^2 \int_{k\neq p_1,p_2} dk\, R_{kk} \,.
\end{align}
Then one can write down the reduced density matrix in perturbative expansion, 
\begin{align}
\rho_A^{\rm (fin)} &= \biggl( u_1^2 +\lambda f +\lambda^2 g
	- \lambda^2 u_1^2\int_{k\neq p_1,p_2} dk\, R_{kk} \biggr) {1 \over V} |{\tilde p}_1\rangle\langle {\tilde p}_1| \nonumber\\
&\quad	+\biggl(u_2^2 -\lambda f -\lambda^2 g 
	- \lambda^2 u_2^2\int_{k\neq p_1,p_2} dk\, R_{kk} \biggr) {1 \over V} |{\tilde p}_2\rangle\langle {\tilde p}_2| \nonumber\\
&\quad	+\lambda^2 \int_{k,k'\neq p_1,p_2}dk dk' R_{kk'} |k\rangle \langle k'| +\CO(\lambda^3) \,, \label{eq:entiniRDM} \\
R_{kk'} &= {1 \over V^2} \int_{l\neq q_1,q_2} dl\, \CT_{kl}\CT_{k'l}^* +\CO(\lambda) \,,
\end{align} 
with
\begin{align}
V^2 f &= 2u_1u_2 (u_1\im\CT_{p_2q_2} -u_2\im\CT_{p_1q_1}) \,, \\
V^4 g &= 4u_1u_2 (u_1 \im\CT_{p_1q_1} +u_2\im\CT_{p_2q_2}) (u_1\im\CT_{p_2q_2}-u_2\im\CT_{p_1q_1})\nonumber\\
&\quad	+u_2^2|\CT_{p_1q_1}|^2 -u_1^2|\CT_{p_2q_2}|^2 
	-{2u_1u_2(u_1+u_2) \over u_1-u_2}\re(\CT_{p_1q_2}\CT_{p_2q_1}) \,. 
\end{align}
Note that $f$ and $g$ are anti-symmetric with respect to the indices 1 and 2. 
Since (\ref{eq:entiniRDM}) implies a reduced density matrix after a similarity transformation, 
we can calculate $S_E^{\rm (fin)}$, the entanglement entropy of the final state. 
Here we introduce $r_k$ $(k\neq p_1,p_2)$ which denotes the eigenvalues of $R_{kk'}$. 
Subtracting the initial entanglement entropy (\ref{eq:entinient}) from $S_E^{\rm (fin)}$, 
we obtain the variation of entanglement entropy as 
\begin{align}
\Delta S_E &= -\lambda^2\log\lambda^2 \int_{k\neq p_1,p_2} dk\,r_k
	-\lambda f \log{u_1^2 \over u_2^2} \nonumber\\
&\quad	+\lambda^2 \biggl( \int_{k\neq p_1,p_2} dk\,r_k(1-S_E^{\rm (ini)}-\log r_k) 
	-{f^2\over 2u_1^2u_2^2} -g\log{u_1^2 \over u_2^2} \biggr)+\CO(\lambda^3) \,. \label{eq:entEEchange}
\end{align}
The leading term is of order $\lambda^2\log\lambda^2$ and is similar to the case of 
the unentangled initial state (\ref{eq:uneEEchange}). 
While the sub-leading term in the case of the unentangled initial state is 
of order $\lambda^2$, 
the sub-leading term in the case of the entangled initial state appears at order $\lambda$. 
This order $\lambda$ contribution comes from the mutual transition between the states, 
$|p_1,q_1\rangle$ and $|p_2,q_2\rangle$. 
When the particles A and B at the initial state are maximally entangled, 
{\it i.e.}, $u_1=u_2 =1/\sqrt{2}$, 
the term of order $\lambda$ vanishes. 

In the same way, one can consider an $n$ coherent state as an initial state, 
namely, $|{\rm ini}\rangle \sim \sum_{j=1}^n u_j |p_j,q_j\rangle$, $\sum_{j=1}^n u_j^2 = 1$. 
Since the final state includes $S_{p_iq_j}V^{-2}|p_iq_j\rangle$, 
we firstly diagonalize the matrix $\mathbb{Q}=(S_{p_iq_j})$ ($i,j=1,\dots,n$) 
so that $W \mathbb{Q} W^{-1} = {\rm diag}(\zeta_1,\dots,\zeta_n)$, 
and replace $|p_i,q_i\rangle$ with $|{\hat p}_i,{\hat q}_i\rangle$ like (\ref{eq:diagvec}).
Then, by a procedure similar to (\ref{eq:defredvec}), 
we can obtain a simplified reduced density matrix like (\ref{eq:entredDM}). 
Therefore the leading contribution to the variation of entanglement entropy is 
$\lambda^2\log\lambda^2 \int_{k\neq p_1,\dots, p_n}dk\int_{l\neq q_1,\dots, q_n}dl\, V^{-2}\CT_{kl}\CT^*_{kl}$, in which $\CT_{kl} = \sum_{j=1}^n u_j\CT_{kl;p_jq_j}$.

\section{Examples}
\label{sec:examp}

\subsection{Field theory with $\phi^4$-like interaction}
\label{sec:phi4}

We consider two real scalar fields, $\phi_A$ and $\phi_B$, of which action 
with a $\phi^4$-like interaction is 
\begin{align}
S = -\int d^{d+1}x \biggl( {1 \over 2}\partial_\mu \phi_A \partial^\mu \phi_A
+{1 \over 2}\partial_\mu \phi_B \partial^\mu \phi_B 
+{1 \over 2}m^2 (\phi_A^2 +\phi_B^2)
+{\lambda \over 4}\phi_A^2 \phi_B^2 \biggr) \,. \label{eq:phi4action}
\end{align}
We focus on a scattering process of 
two incoming particles and two outgoing particles such as ${\rm A}+{\rm B} \to {\rm A}+{\rm B}$.
Since we can assume that the incoming and outgoing particles are free on-shell particles 
in the far past and future, 
one can describe a Fock space of such (1+1)-particle states as 
\begin{align}
|\vp,\vq\rangle = a_\vp^\dagger|0\rangle_A \otimes b_\vq^\dagger |0\rangle_B \,. 
\end{align}
$a_\vp^\dagger$ and $b_\vq^\dagger$ are the creation operators of particles A and B 
and are defined by the following mode expansion for free scalar fields:
\begin{align}
\phi_A(x) = \int {d^d\vp \over (2\pi)^d} {1 \over 2E_\vp}(a_\vp e^{-i p \cdot x} 
	+a_\vp^\dagger e^{i p \cdot x}) \,, \quad
\phi_B(x) = \int {d^d\vq \over (2\pi)^d} {1 \over 2E_\vq}(b_\vq e^{-i q \cdot x} 
	+b_\vq^\dagger e^{i q \cdot x}) \,,
\end{align}
where $p^0 = E_\vp = \sqrt{\vp^2 + m^2}$. 
The factor $d^d\vp/((2\pi)^d 2E_\vp)$ is a Lorentz invariant integration measure. 
The creation and annihilation operators obey the commutation relations:
\begin{align}
[a_\vp,a_\vk^\dagger] = 2E_\vp (2\pi)^d \delta^{(d)}(\vp-\vk) \,, \quad 
[b_\vq,b_\vl^\dagger] = 2E_\vq (2\pi)^d \delta^{(d)}(\vq-\vl) \,. 
\end{align}

Now let us study the case that the initial state is $|{\rm ini}\rangle = |{\vec p_1},{\vec q_1}\rangle$. 
Since the identity operator on the (1+1)-particle Hilbert space is 
\begin{align}
({\bf 1})_\text{(1+1)-particle} = \int {d^d\vp \over (2\pi)^d}{1 \over 2E_\vp} {d^d\vq \over (2\pi)^d} {1 \over 2E_\vq} |\vp,\vq\rangle \langle \vp,\vq| \,, 
\end{align} 
the final state (\ref{eq:finsta}) is described as 
\begin{align}
&\quad|{\rm fin}\rangle = {\bf S}|{\rm ini}\rangle 
= \int {d^d\vk \over (2\pi)^d}{1 \over 2E_\vk} {d^d\vl \over (2\pi)^d} {1 \over 2E_\vl} |\vk,\vl\rangle \langle \vk,\vl|{\bf S}|{\vec p_1},{\vec q_1}\rangle \nonumber\\
&= {1 \over 2E_{\vec p_1} 2E_{\vec q_1} L^{2d}} |{\vec p_1},{\vec q_1}\rangle \langle {\vec p_1},{\vec q_1}|{\bf S}|{\vec p_1},{\vec q_1}\rangle 
+ \int_{\vk\neq{\vec p_1} \atop \vl\neq{\vec q_1}} {d^d\vk \over (2\pi)^d}{1 \over 2E_\vk} {d^d\vl \over (2\pi)^d} {1 \over 2E_\vl} |\vk,\vl\rangle \langle \vk,\vl|i{\bf T}|{\vec p_1},{\vec q_1}\rangle \,, \label{eq:phi4fin}
\end{align}
where $L$ originates from the spacial volume of phase space, $L^d = (2\pi)^d \delta^{(d)}(0) = \int d^d \vx\, e^{i\vx\cdot {\vec 0}}$. 
The final state (\ref{eq:phi4fin}) does not contain the states proportional to 
$|\vk(\neq {\vec p_1}),{\vec q_1}\rangle$ and $|{\vec p_1},\vl(\neq {\vec q_1})\rangle$, 
which appear in the second line of (\ref{eq:unefin}), 
because such states vanish due to the factor of momentum conservation in the S-matrix element, 
namely, $\langle \vk,\vl|{\bf S}|{\vec p_1},{\vec q_1}\rangle \sim \delta^{(d+1)}(k+l-p_1-q_1)$. 

As we have studied in Section \ref{sec:pertu}, 
the variation of entanglement entropy in a scattering process is determined 
by the transition matrix ${\bf T}$. 
From the action (\ref{eq:phi4action}), we perturbatively calculate the S-matrix element, 
\begin{align}
\langle \vk,\vl|{\bf S}|{\vec p_1},{\vec q_1}\rangle 
&= 2E_{\vec p_1} 2E_{\vec q_1} (2\pi)^d \delta^{(d)}(\vk - {\vec p_1}) (2\pi)^d \delta^{(d)}(\vl - {\vec q_1}) \nonumber\\
&\quad	-i\lambda (2\pi)^{d+1}\delta^{(d+1)}(k+l-p_1-q_1) +\CO(\lambda^2) \,. 
\end{align}
Substituting this S-matrix element into (\ref{eq:phi4fin}), we obtain the final state. 
Then the reduced density matrix automatically becomes block-diagonal, 
\begin{align}
\rho_A^{\rm (fin)} &= \biggl(1 -\lambda^2 
	\int_{\vk \neq {\vec p_1}} {d^d\vk \over (2\pi)^d}{1 \over 2E_\vk} M_{\vk\vk} \biggr)
	{1 \over 2 E_{\vec p_1}L^d}|{\vec p_1}\rangle \langle {\vec p_1}| \nonumber\\
&\quad +\lambda^2 
	\int_{\vk \neq {\vec p_1}} {d^d\vk \over (2\pi)^d}{1 \over 2E_\vk} 
	M_{\vk\vk} {1 \over 2E_\vk L^d} |\vk\rangle \langle\vk| +\CO(\lambda^4) \,, \\
M_{\vk\vk} &= {1 \over 2 E_{\vec p_1} 2 E_{\vec q_1} 2E_{{\vec p_1}+{\vec q_1} -\vk} L^d }
	\bigl\{ 2\pi\delta(E_\vk +E_{{\vec p_1}+{\vec q_1} -\vk} -E_{\vec p_1} -E_{\vec q_1})\bigr\}^2 \,. 
\end{align}
Notice that we have normalized this density matrix so that $\tr_A \rho_A^{\rm (fin)} = 1$. 
Then the variation of entanglement entropy (\ref{eq:uneEEchange}) is computed as 
\begin{align}
\Delta S_E &= -\lambda^2\log\lambda^2 
	\int_{\vk \neq {\vec p_1}} {d^d\vk \over (2\pi)^d}{1 \over 2E_\vk} M_{\vk\vk} \nonumber\\ 
&\quad	+\lambda^2 \int_{\vk \neq {\vec p_1}} {d^d\vk \over (2\pi)^d}{1 \over 2E_\vk} 
	M_{\vk\vk} \biggl(1 -\log {M_{\vk\vk} \over 2E_\vk L^d}\biggr) + \CO(\lambda^4) \,. \label{eq:entphi4}
\end{align}
We shall calculate it further by employing a center of mass frame, 
that is, ${\vec p_1} = - {\vec q_1} =: {\vec p}_{cm}$ and $\sqrt{{\vec p_{cm}}{}^2 +m^2} =: E_{cm}$. 
Of course the momenta of outgoing particles obey $\vk = -\vl$ due to the momentum conservation. 
Then the $d$-dimensional integration can be replaced with a spherical integration 
as $d^d \vk = dk d\Omega_{d-1} k^{d-1}$, 
because the integration kernel in (\ref{eq:entphi4}) depends only on the norm of $\vk$.
Therefore we finally obtain 
\begin{align}
\Delta S_E &=  -\lambda^2\log\lambda^2 {\pi^{1-{d \over 2}} \over 2^{d+3} \Gamma({d \over 2}) L^{d-1}} 
	{|{\vec p}_{cm}|^{d-2} \over E_{cm}^3} \nonumber\\
&\quad	+\lambda^2 {\pi^{1-{d \over 2}} \over 2^{d+3} \Gamma({d \over 2}) L^{d-1}} 
	{|{\vec p}_{cm}|^{d-2} \over E_{cm}^3} \bigl( 1+\log(16E_{cm}^4 L^{2d-2}) \bigr)
	+\CO(\lambda^4) \,. \label{eq:ftEEcm}
\end{align}
When the number of the spacial dimension $d$ is equal to three, the leading term of the variation of entanglement entropy is proportional 
to $|{\vec p}_{cm}|/E_{cm}^3$. 
This is consistent with the cross section, which is 
$(d\sigma / d\Omega)_{cm} =  {\lambda^2 \over 64\pi^2 } |{\vec p}_{cm}| / E_{cm}^3$, 
because both the variation of entanglement entropy and the cross section 
originate from a square of the absolute value of the scattering amplitude. 
Notice that the remaining factor $L$ in the entanglement entropy is an artifact 
caused by choosing the single-mode initial state whose norm has delta-functional divergence. 
The volume dependence of entanglement entropy in field theories was discussed 
also in Ref.~\cite{Balasubramanian:2011wt,Hsu:2012gk}, 
where the momentum-space entanglement entropy is proportional to a spacial volume. 
The difference between the volume dependence of Ref.~\cite{Balasubramanian:2011wt,Hsu:2012gk} and ours 
is mostly caused by the absence of integration 
with respect to the initial state momenta in our calculation.

\subsection{Time-dependent interaction in quantum mechanics}
\label{sec:timedep}

In this subsection we turn to quantum mechanics 
with a time-dependent interaction, $\lambda H_{\rm int}(t)$. 
We set the initial state so that $|{\rm ini}\rangle = |p_1,q_1\rangle$ at $t=0$. 
Then the time evolution of this initial state is described as 
\begin{align}
|\Psi(t){\rangle} &= |p_1,q_1{\rangle} 
	+\lambda\sum_{k \neq p_1}C_{k q_1;p_1q_1}(t)e^{-i E_k t}|k,q_1 \rangle 
	+\lambda\sum_{l \neq q_2}C_{p_1 l;p_1q_1}(t)e^{-i E_l t}|p_1,l \rangle \nonumber\\
&\quad	 +\lambda\sum_{k\neq p_1,l\neq q_1} C_{k l;p_1q_1}(t)e^{-i E_{kl} t}|k,l\rangle  \,, \label{tdepfin} 
\end{align}
up to normalization. 
$E_p$, $E_q$ and $E_{pq}$ are energy eigenvalues 
which are defined in terms of the non-interacting part of the Hamiltonian (see (\ref{eq:Hami})), 
\begin{align}
H_A |p\rangle_A = E_p |p\rangle_A \,, \quad 
H_B |q\rangle_B = E_q |q\rangle_B \,, \quad 
H_0 |p,q\rangle = E_{pq} |p,q\rangle \,. 
\end{align}
The interacting Hamiltonian $\lambda H_{\rm int}(t)$ yields $C_{kl;pq}(t)$. 
By the use of the well-known time-dependent perturbation theory, 
we can calculate 
\begin{align}
C_{kl;p_1q_1}(t) = -i\int_0^t dt' e^{i\omega_{kl;p_1q_1}t'}T_{kl;p_1q_1}(t') \,, \quad 
T_{kl;p_1q_1}(t) := \langle k,l|H_{\rm int}(t)|p_1,q_1\rangle \,. 
\end{align}
where $\omega_{kl;p_1q_1} := E_{kl} -E_{pq}$. 
Since the time-dependent density matrix of $|\Psi(t)\rangle$ is given by 
$\rho(t) = |\Psi(t)\rangle \langle \Psi(t)|$, 
we can calculate the reduced density matrix $\rho_A(t) = \CN^{-1} \tr_B \rho(t)$ 
together with the normalization by $\tr_A\rho_A = 1$. 
After the same procedure as Ref.~\cite{Balasubramanian:2011wt} or Section \ref{subsec:unent}, 
we obtain the entanglement entropy, 
\begin{align}
S_E(t) = -\lambda^2\log\lambda^2 \sum_{k\neq p_1,l\neq q_1} 
	\int_0^t dt' \int_0^t dt''\, e^{i\omega_{kl;p_1q_1}(t'-t'')} T_{kl;p_1q_1}(t')   T_{kl;p_1q_1}^*(t'') +\CO(\lambda^2) \,. 
\end{align}
Notice that one can regard $\lambda T_{kl;p_1q_1}(t=\infty)$ as a kind of transition matrix.

\section{Conclusion and discussion}
\label{sec:concl}

We have studied the variation of entanglement entropy from an initial state to a final state 
in a scattering process. 
We concentrated on the scattering of $2 \to 2$ particles and 
perturbatively calculated the entanglement entropy of final states 
for the two kinds of simple initial states: 
the unentangled state (\ref{eq:uneini}) and the entangled state (\ref{eq:entini}). 
In both cases the leading terms of the variation of entanglement entropy, 
(\ref{eq:uneEEchange}) and (\ref{eq:entEEchange}), are of order $\lambda^2\log\lambda^2$ 
and are proportional to the trace of a square of the absolute value of T-matrix elements, 
which are, in other words, the scattering amplitudes. 
The next leading term in the case of the unentangled initial state is of order $\lambda^2$. 
On the other hand the next leading term in the case of the entangled initial state appears 
at order $\lambda$, because there is a mutual transition 
between the states $|p_1,q_1\rangle$ and $|p_2,q_2\rangle$. 

We have considered the model of two real scalar fields with the $\phi^4$-like interaction 
as an example in a field theory. 
The variation of entanglement entropy has been computed perturbatively. 
If we employ the center of mass frame, the leading term (at order $\lambda^2\log\lambda^2$) in 
the variation of entanglement entropy depends on the momenta of 
initial particles as $|\vec p_{cm}|^{d-2}/E_{cm}^3$. 
Notice that this factor becomes $|\vec p_{cm}|/E_{cm}^3$, 
when the space dimension is equal to three, {\it i.e.}, the coupling $\lambda$ is dimensionless. 
The same factor also appears in the cross section, because it originally comes 
from the scattering amplitude. 
Therefore, as we expected, the variation of entanglement entropy is proportional to 
the cross section. 

We have also mentioned the time-dependent interaction as an example in quantum mechanics. 
The time-evolution of entanglement entropy from the simple initial state $|p_1,q_1\rangle$ can be 
written in terms of the transition matrix at the leading order $\lambda^2\log\lambda^2$. 

With the AdS/CFT correspondence, one can identify the scattering amplitude 
in a field theory of strong coupling with $\exp(-\CA)$, 
where $\CA$ is an area of minimal surface in a bulk gravity theory, 
while the holographic entanglement entropy \cite{Ryu:2006bv,Ryu:2006ef} is given by $\CA'/(4 G_N)$, 
where $\CA'$ is an area of another minimal surface. 
That is to say, both of the scattering amplitude and entanglement entropy 
in a strongly coupled field theory are associated with minimal surfaces from the point 
of view of the AdS/CFT correspondence. 
In this paper we have shown the relation 
between the scattering and the variation of entanglement entropy 
by the perturbative calculations in a weak coupling. 
It is then in order to ask whether we can clarify such a relation 
from a field theory in a strong coupling. 
For this purpose we need to test it in an exactly calculable model. 
Moreover, the holographic understanding of such a relation, 
or a relation between those minimal surfaces, is another problem for the future.


\acknowledgments

This work was supported by Mid-career Researcher Program 
through the National Research Foundation of Korea (NRF) grant No.~NRF-2013R1A2A2A05004846. 
IP thanks Sang-Jin Sin for his hospitality during the IP's visit to Hanyang University through its foreign scholar invitation program. 
SS was also supported in part by Basic Science Research Program through NRF grant No.~NRF-2013R1A1A2059434.

\end{document}